\documentclass[conference,10pt]{IEEEtran}

\usepackage{amsmath}
\usepackage{amssymb}
\usepackage{amsfonts}
\usepackage{graphicx}
\usepackage{enumerate}

\usepackage{mathabx} % Added by Babis
\usepackage{flushend}
\usepackage{cite}
\usepackage{subfigure}
\usepackage{url}
\usepackage{array}
\usepackage{verbatim}
\usepackage{comment}
\usepackage{stfloats}

\usepackage{color}
\definecolor{red}{rgb}{1,0,0}
\definecolor{blue}{rgb}{0,0,1}

\long\def\comment#1{}

\newfont{\bbb}{msbm10 scaled 700}

\newfont{\bb}{msbm10 scaled 1100}

\newcommand{\EE}{\mbox{\bb E}}

\newcommand{\av}{{\bf a}}
\newcommand{\bv}{{\bf b}}

\newcommand{\hv}{{\bf h}}

\newcommand{\wv}{{\bf w}}
\newcommand{\vv}{{\bf v}}
\newcommand{\xv}{{\bf x}}
\newcommand{\yv}{{\bf y}}

\newcommand{\Id}{{\bf I}}

\newcommand{\Wm}{{\bf W}}

\newcommand{\Ym}{{\bf Y}}

\newcommand{\Ac}{{\cal A}}

\newcommand{\Cc}{{\cal C}}

\newcommand{\Ec}{{\cal E}}

\newcommand{\Kc}{{\cal K}}
\newcommand{\Lc}{{\cal L}}

\newcommand{\Nc}{{\cal N}}

\newcommand{\Sc}{{\cal S}}

\newcommand{\transp}{{\sf T}}

\def\argmax{\mathop{\rm arg\,max}}

\begin{document}
\title{ RRH based Massive MIMO \\ with ``on the Fly'' Pilot Contamination Control}
\author{\IEEEauthorblockN{Ozgun Y. Bursalioglu, Chenwei Wang, Haralabos Papadopoulos}
\IEEEauthorblockA{Wireless Systems Project, MNTG\\
Docomo Innovations Inc. Palo Alto, CA, 94304, USA\\
 obursalioglu, cwang, hpapadopoulos@docomoinnovations.com\\
} %\vspace*{-0.25in}
\and
\IEEEauthorblockN{Giuseppe Caire}
\IEEEauthorblockA{Communications and Information Theory Group, \\
Technische Universit{\"a}t Berlin, 10587 Berlin, Germany\\
caire@win.tu-berlin.de} %\vspace*{-0.25in}
}

\maketitle
%\vspace{-5cm}
\begin{abstract}
\vspace{-0cm}
 Dense large-scale antenna deployments are one of the most promising technologies for delivering very large  throughputs per unit area in the downlink (DL) of cellular networks.  We consider such a dense deployment involving a distributed system formed by multi-antenna remote radio head (RRH) units connected to the same fronthaul serving a geographical area. Knowledge of the DL channel between each active user and its nearby  RRH antennas is most efficiently obtained at the RRHs via reciprocity based training, that is, by estimating a user's channel using uplink (UL) pilots  transmitted by the user, and exploiting the UL/DL channel reciprocity.  

We consider aggressive pilot reuse across an RRH system, whereby  a single pilot dimension is simultaneously assigned to multiple active users. We introduce a novel coded pilot approach, which allows each RRH unit to detect pilot collisions, i.e., when more than a single user in its proximity uses the same pilot dimensions. Thanks to the proposed coded pilot approach, pilot contamination can be substantially avoided. As shown, such strategy can yield densification benefits in the form of increased multiplexing gain per UL pilot dimension with respect to conventional reuse schemes and some recent approaches assigning pseudorandom pilot vectors to the active users.

\end{abstract}

%\newpage


%%%%%%%%%%%%%%%%%%%%%%%%%%%%%%%%%%%%%%%%%%
\vspace{-0.2cm}
\section{Introduction}
\label{section:introduction}
\vspace{-0.15cm}
Dense large-scale MIMO deployments are an attractive option for providing the vast throughputs per unit area needed to cope with the explosive growth in wireless traffic.
Small cells \cite{andrews-femtocell-survey}  enable  dense spatial resource reuse, i.e.,  coexistence of spatially separated short-range links on the same channel resource.  
Combined with large antenna arrays to spatially multiplex many users on the same channel resource \cite{marzetta-massive,Huh11}, dense deployments can potentially provide 100-fold or higher increases in throughput per unit area and bandwidth. Such dense massive MIMO operation is possible  at higher frequencies (e.g., 6-60 GHz), where large numbers of antennas can be packed in a  small form factor, \cite{adhikary-mmWave-jsac}, \cite{rappaport-mmWave}. 

In order to achieve large spectral efficiencies in the downlink (DL) via multiuser (MU) MIMO, 
channel state information at the transmitter (CSIT) is needed. Following the massive MIMO approach  \cite{marzetta-massive},  
CSIT can be obtained  from the users' uplink (UL) pilots via Time-Division Duplexing (TDD) and  UL/DL radio-channel reciprocity.  This  allows training large antenna arrays by allocating as few UL pilot dimensions as the number of single-antenna users simultaneously served.

Although from the point of view of training a massive array at a single site the pilot efficiency of reciprocity-based training  is very attractive, to enable operation in a dense antenna-site environment the uplink pilot dimensions need to be aggressively  reused. However, having nearby users transmit the same pilots  can lead to significant pilot contamination at nearby sites and can  greatly impact performance.

In \cite{marzetta-massive}, for example, a macro-cellular network is considered and spatial pilot-reuse of 7 is advocated to alleviate pilot contamination. Such a large pilot-reuse distance, however, is equivalent to a very poor spatial reuse of resources. In \cite{Huh11} geographical scheduling across the cellular network is exploited to optimize the spatial reuse and the MIMO method separately at cell-center and cell-edge locations throughout the cellular layout. 
%Geographical scheduling amounts to coordinated scheduling among BSs, so that in each slot (or resource block) each BS in the network is restricted to schedule users in specific geographic (relative to the BS) location.
 As a result, high spectral efficiencies can be achieved with reuse-one pilot assignments  to cell-center users, while reuse-3 can be exploited at the cell-edge. 
 %Both \cite{marzetta-massive}, and \cite{Huh11} assume regular macro layouts and are not ideally suited for massive MIMO small cells. Regarding \cite{marzetta-massive},  irregular and dense small cell layouts make the situation a lot worse and the densification is not fully realized. Similarly,  \cite{Huh11}  relies on a ``symmetric'' layout and uniform load per cell, which are clearly not the case with dense irregular small-cell deployments. Furthermore it relies on coordinated scheduling across the BS.
Another line of work to avoid pilot contamination includes exploiting the knowledge of covariance matrices to allocate pilot resources to users based on their  support of angle of arrival \cite{Gesbert-aoa}. Unlike \cite{Gesbert-aoa}, we schedule users randomly.

Pilot assignment in dense antenna-site deployments  is much more challenging. First, due to the typically irregular antenna-site layouts different user terminals may train different numbers of nearby antennas. Unlike the symmetric macro scenario considered in \cite{Huh11}, there are no simple geographic rules that result in scheduling users across the network with symmetric pilot-contamination characteristics, thereby making the problem of
optimized coordinated scheduling and pilot assignments across the network a non-trivial one. 

%{\RED In \cite{Larsson-Marzetta} an   system is considered according to which each active user in the network is assigned a pseudo-random pilot vector...}

In this work, we consider aggressive reuse of the pilot dimensions across a remote radio head (RRH) system.  The combination of aggressive pilot reuse and random user scheduling inherently results in pilot contamination and pilot collisions at different RRH sites. By assigning the same pilot dimension to multiple users across the RRH coverage area for simultaneous UL pilot transmission, and by employing fast user proximity detection at each RRH site based on these transmissions, different RRH sites can  serve the packets of multiple users  whose codes are aligned on the same pilot dimension. As a result, densification benefits can be achieved and the multiplexing gain of the system can be substantially increased compared to traditional schemes. 

A distributed massive MIMO system with single antenna at each location is considered in \cite{Larsson-Marzetta}, whereby multiple users broadcast pilots over the same pilot dimensions causing pilot contamination. \cite{Larsson-Marzetta} proposes a greedy algorithm for pilot code design and a power allocation optimization between each antenna and user to mitigate pilot contamination. Our work is different from \cite{Larsson-Marzetta} in that it relies on pilot allignment, and fast user proximity detection at each fast RRH site (which can also be viewed as a decentralized RRH-site selection method for each user's packet). More important, unlike \cite{Larsson-Marzetta}, we also advocate the use of large antenna arrays at each RRH site as a means for reducing the number of RRH sites needed to achieve a certain multiplexing gain. As we demonstrate, by leveraging the inherently narrow angular spread in the user channels, large antenna arrays at each RRH site, aggressive pilot reuse, and fast user RRH-sector proximity detection,   large increases in multiplexing gains can be harvested at a fraction of the RRH sites required by single antenna RRH deployments such as \cite{Larsson-Marzetta}.

\section{System Model}
\label{sec_system_model}
We consider a setting involving an RRH system  comprised of  $N$  $M$-antenna radio heads uniformly (and randomly) distributed over a square wrap-around geographical region $\Ac$ with area $A$.  The RRH system serves a large set $\Kc_{\rm tot}$ of user terminals (uniformly and randomly distributed over the RRH coverage region)  via reciprocity-based MIMO over OFDM. 
%We assume slotted transmission according to which assume that the OFDM plane is partitioned into resource blocks over which each user channel is quasi-static.  

We assume a slotted system according to which the  RRH system schedules users for transmission over scheduling slots.  Each slot comprises a subset of concurrent resource blocks (RBs), with each RB corresponding to a contiguous block of OFDM resource elements (REs). Without loss of generality we consider a quasistatic channel model where the user-channels remain fixed within any RB, but are independent across RBs. 

We  consider a  generic scheduling slot $t$, and assume that the users with indices  from $\Kc(t)\subset\Kc_{\rm tot}$ are active in this slot for some preselected scheduling size $K=|\Kc(t)|$. We let $L$ denote the number of RBs in the slot and $Q$ the number of dimensions (REs)  allocated  for uplink pilots in each RB. 
The $k$-th active user (for any given $k\in \Kc(t)$) broadcasts a $Q\times 1$ uplink pilot in pilot RE $n$ given by  $\sqrt{\gamma_{\rm p}}\av_{k}[n]$, where $\av_{k}[n]$ denotes the unit-norm normalized version of the UL pilot vector assigned to the user by the RRH system and where $\gamma_{\rm p}$ represents the a priori known UL pilot transmit energy.

The received signal at the $M$-dimensional array of RRH site $j$ from all pilot transmissions during the $Q$  pilots REs  on RB $n$ can be expressed (after rescaling by $1/\sqrt{\gamma_{\rm p}}$) in the form of the following $q\times M$ matrix (The dependence on $t$ for $\Ym_j[n], \,\xv_j[n], $ etc. is suppressed in (\ref{Ymj-pilots})--(\ref{received-signal-userk}).)
\begin{equation}
\label{Ymj-pilots}
\Ym_j[n] = \sum_{k=\Kc(t)} \av_{k}[n] \hv^\transp_{kj}[n]  +\Wm_j[n]
\end{equation}
where $\hv_{kj}[n]\sim \Cc\Nc(0,g_{kj}\Id)$ denotes the  channel between the antenna of the $k$-th active user and the $M$ antennas of RRH site  $j$. The $\hv_{kj}[n]$'s are independent in $k,j,n$. We assume that RRH site $j$ does not know a priori the $g_{kj}$'s. $\Wm_j[n]$ represents noise compromising of IID $\Cc\Nc(0,N_o/\gamma_{\rm p})$ entries, where $N_o$ denotes the thermal noise power.

In the system we consider RRH $j$ has available for (potential) transmission to user $k$ in RB $n$ a coded packet $u_k[n]$ (common across all RRHs). We focus on linear precoding options whereby, during the data transmission portion of the RB $n$,
RRH site $j$ transmits the following $M\times 1$ vector signal over its $M$-dimensional array
\begin{equation}
\label{precoding-signal-BSj}
\xv_j[n] = \sum_{k\in\Sc_j(t)} \vv_{k}[n] u_{k}[n]
\end{equation}
where   $\vv_{k}[n]$ denotes the  precoding vector for user $k$ and $\Sc_j(t)\subset\Kc(t)$ is a suitably chosen subset of active users.
The set  $\Sc_j(t)$ for which RRH $j$ transmits their packet at slot $t$ and the precoding vectors  $\{\vv_k\}$  are chosen based on the received signal over the $Q$ UL pilot REs in RB $n$.  The received signal at active user $k$ during the data-transmission portion is  
\begin{equation}
\label{received-signal-userk}
y^{\rm DL}_k[n] = \sum_{j=1}^N  \hv^\transp_{kj}[n] \xv_{j} [n] +w^{\rm DL}_k[n] 
\end{equation}
where $w^{\rm DL}_k[n]\sim \Cc\Nc(0,N_o)$ represents thermal noise.  

In general, for a RRH system with a sufficiently large coverage area, each user pilot is received at  ``sufficiently'' high power by  only a fraction of RRH sites  in the proximity of the users, i.e., only by RRH sites with sufficiently large $g_{kj}$'s.  For simplicity we consider a distance-based user RRH-site  proximity model, according to which a user pilot is received at  ``sufficiently'' high power by RRH site $j$ if the distance between the user and the RRH site is less than  $r_o$, for some value $r_o$. As a result, a user can be served by only the RRH sites within a distance $r_o$ from the user. Given that a user  can also be interfered by RRH sites within a distance $r_o$, we assume that a user can be served if and only if the user experiences  {\em no} pilot contamination by any RRH site within distance $r_o$ to the user. 

This modeling abstraction is reasonable for reciprocity-based DL MIMO transmission (as the pilot contamination from a RRH site to  a user depends on the large-scale  channel strength between the RRH and the user \cite{marzetta-massive}) and corresponds to  neglecting pilot contamination from RRH sites at distances larger than $r_o$.  It is especially justified for milimeter Wave (mmWave) channels, where the blocking probability grows such rapidly with distance that it is reasonable to assume that beyond a certain distance no signal is received, despite the purely distance-based pathloss which, although sharply decreasing function of distance, may be nevertheless non-zero. 
Letting $\Kc_j(t)\subset \Kc(t)$ denote the subset of  active users in proximity of RRH $j$ in slot $t$,  the set of active users served by RRH $j$ must thus satisfy $\Sc_j(t)\subset \Kc_j(t)$.

We  focus on pilot schemes where the $Q$ pilot REs in an RB are split into disjoint groups of $q$ pilot dimensions (there are $Q/q$ such groups). When $q>1$, the users sharing a group of $q$ pilot REs are assigned pseudorandomly generated codewords. 

\begin{figure}
\centering
\centerline{\includegraphics[width=4cm]{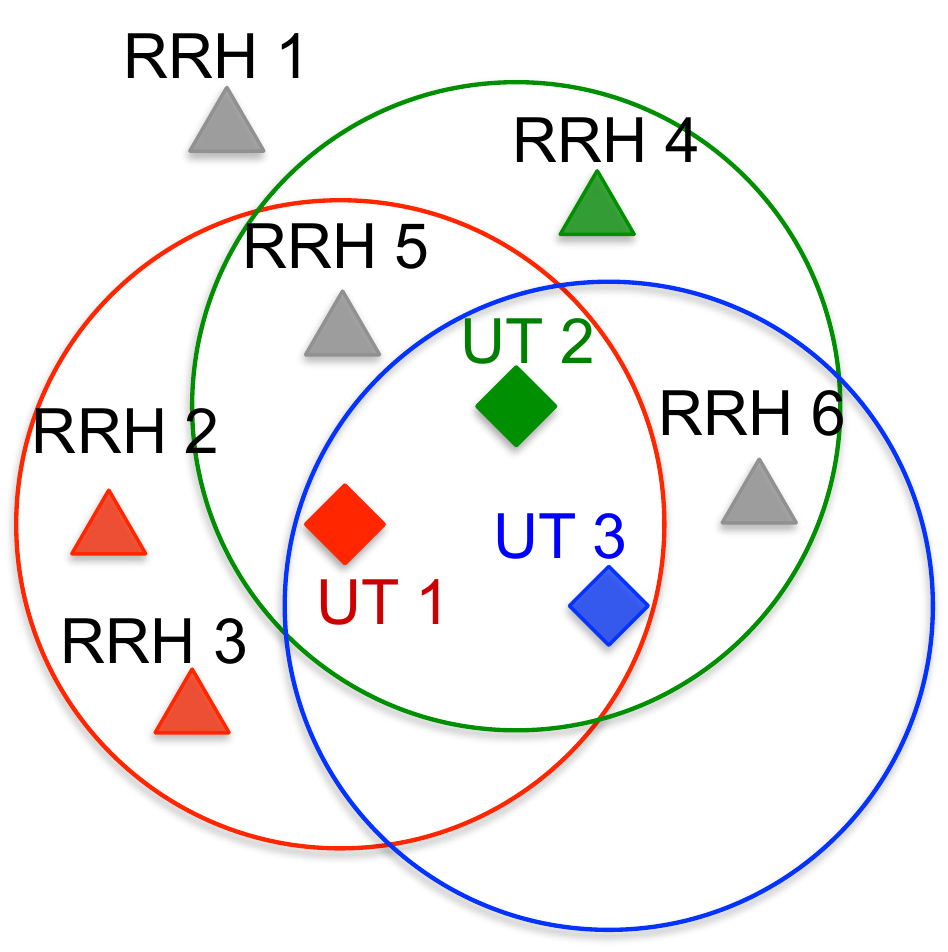}}
\vspace*{-.1in}
\caption{$q=Q=1$, 3 active user terminals (UTs) and 6 RRH sites.}
\label{fig_rrh_reqs}
\vspace*{-.2in}
\end{figure}

The scenario is illustrated via the toy example in Fig.~\ref{fig_rrh_reqs} involving $q=Q=1$, an RRH system with 6 RRH sites, serving  3 active user terminals (UTs). The 3 UTs broadcast pilots on the same pilot RE on an RB in slot $t$. As it can be seen in the figure, RRH 1 can serve none of the UTs as it is not in the proximity of any of the  UTs. In contrast, RRH sites  2 and 3 are in proximity of only  UT 1 and transmit the same coded packet $u_1[n]$ to UT 1. Similarly,  RRH 4 transmits $u_2[n]$ to  UT 2. In contrast, RRH sites 5 and 6 are in the vicinity of multiple UTs (pilot collision event) and thus serve {\em none} of the UTs. It is also evident that UT 3 is not served in the given scheduling slot as its transmitted pilot is contaminated (collided) at each RRH in its proximity by other user terminals. Then, $\Sc_2(t)=\Sc_3(t)=\{1\}$, $\Sc_4(t)=\{2\}$, and $\Sc_1(t)=\Sc_5(t)=\Sc_6(t) = \emptyset$. In summary,  three active UTs broadcast pilots on a common pilot RE, and the 6 RRH-site  system can serve two of these UTs yielding an instantaneous multiplexing gain equal to 2. 

We consider a user $k$ as  ``served'' by the RRH system at $t$, if its packet is transmitted by {\em at least one} RRH site in its vicinity, i.e., if and only if $\exists j$ s.t., $k\in \Sc_j(t)$.  Then, letting $\Sc(t)=\cup_j\Sc_j(t)$, the multiplexing gain of the RRH system in slot $t$ is given by $|\Sc(t)|$. An implicit assumption in calling this the RRH-system instantaneous multiplexing gain is that for any user $k\in \Sc(t)$, any RRH $j'$ within distance $r_o$  must also {\em not create} pilot contamination at user $k$. When user codes are aligned on a single pilot RE then no RRH serves active users in a pilot dimension when multiple active users in the pilot dimension are in the proximity of the RRH. 

Similarly, consider the case where a set of active users share  $Q = q>1$ pilot REs on an RB and assume the users are assigned pseudorandom pilots over the $q$ pilot REs on an RB so that the pilots of any $q$ active users are linearly independent. In the same spirit as in the $q=1$ case, the RRH serves all the active UTs (on the shared $q$ pilot REs) in proximity of the RRH, if no more than $q$ UTs  are in the proximity of the RRH, and serves no UTs otherwise.  Then for $q=Q$,
\vspace{-0.2cm}
\begin{equation}
\label{Sjtq}
\Sc_j(t) = \Sc_j(t;q) =
\begin{cases}
\emptyset & \text{if $|\Kc_j(t)|>q$}\\
\Kc_j(t) & \text{if $|\Kc_j(t)|\leq q$}\
\end{cases}
\end{equation}
\vspace{-0.6cm}

\section{Multiplexing Gains with Genie-Aided Proximity Detection }
\vspace{-0.15cm}
We next consider the genie-aided scenario according to which each RRH {\em knows} the identities and the pilot codes of the active user terminals that are in its proximity. A method for achieving such knowledge based on coded UL pilots is presented in Sec.~\ref{sec_codes}. We  investigate the average multiplexing gains per pilot RE that can be obtained over coverage area $\Ac$ during a sufficiently large number of scheduling slots, $T$. Given that the multiplexing gains per dimension for the $Q/q = 1$ setting are the same as those for the $Q/q>1$ setting it suffices to study the multiplexing gains per RE in the case $q=Q$: 
\begin{equation}\label{mq}
m_{q}(K,N) = \lim_{T\to \infty}\frac{1}{Tq}\sum_{t = 1}^T |\Sc(t;q)|,
\end{equation}
with $\Sc(t;q) = \cup_j\Sc_j(t;q)$.
The maximum multiplexing gain per pilot RE for a given $q=Q$ scheme is given by
\begin{equation}
\label{mq-opt}
m_q^*(N) =\max_K m_{q}(K,N)
\end{equation}
with the optimizing active-user scheduling size given by

\begin{equation*}
K^*_q(N) = \argmax_K m_{q}(K,N)
\end{equation*}
\vspace{-0.5cm}
%Finally we let $m^{*\rm total}_{Q,q}(N)$ denote the total average multiplexing gains that can be achieved by $Q$ many pilot dimensions in total and $q$ many pilot dimensions per group for a given $N$ and optimized $K$ values. 

%, and let $\Kc_j$ denote the subset of  active users in proximity of RRH $j$. RRH $j$ serves the packets  of $\Sc_j\subset \Kc_j$ users.  $|\Sc_j|$  may take values from $0$ to $\min\{Q, \, \Kc_j\}$, depending on the number of users in proximity of RRH $j$, and the pilot codes of the users in $\Kc_j$.  
%
%To see this consider the case $Q=1$ according to which a single pilot dimension is allocated for pilot transmission in each resource block.  Clearly the RRH can at most serve one user per resource block. It is also clear in this case that $\Sc_j=\empty$ if $\Kc_j=\empty$ (no user in proximity), but also if $|\Kc_j|>1$ (which corresponds to a pilot collision). To maximize the RRH multiplexing gains  the RRH $j$ will use $\Sc_j=\Kc_j$ in the case that a single user is in proximity i.e., $|\Kc_j|=1$.  Alternatively, consider the case $Q>1$  and where  all $K$ active users are given pseudorandomly generated pilot codewords, so that any $Q$ pilots are a linearly independent set. In this case if  $|\Kc_j|>Q$ we have the event of pilot collision so that $\Sc_j=\empty$. At the same time if $|\Kc_j|\le Q$, the RRH can estimate the channels of the user set $\Kc_j$. Hence, to maximize multiplexing gains, the RRH BS transmits from RRH $j$ the packets of user set  $\Sc_j = \Kc_j$.

\subsection{Upper Bounds based on Structured Scheduling}

Upper bounds on the multiplexing gain per pilot RE can be obtained by assuming that the region $\Ac$ is blanketed with infinitely many RRH sites and users and assuming the ability to freely schedule users on suitably chosen locations. For this upper bound we focus on $q =Q= 1$. On a given slot, our aim is to schedule in $\Ac$ as many as users possible that can be served by an RRH without causing pilot contamination to other users, thereby obtaining an upper-bound on the multiplexing gains per pilot RE with  randomly scheduled users and randomly placed RRHs. Since the area is completely covered by RRHs, a scheduled user can be served as long as it has an infinitesimal area in its disc of radius $r_0$ with no other user disc overlaping. This can be achieved by packing as many discs as possible over the coverage area $\Ac$ with a non-overlapped area per  disc. 

\begin{figure}
\centering
\centerline{\hspace{-0cm}\includegraphics[width=3.1cm]{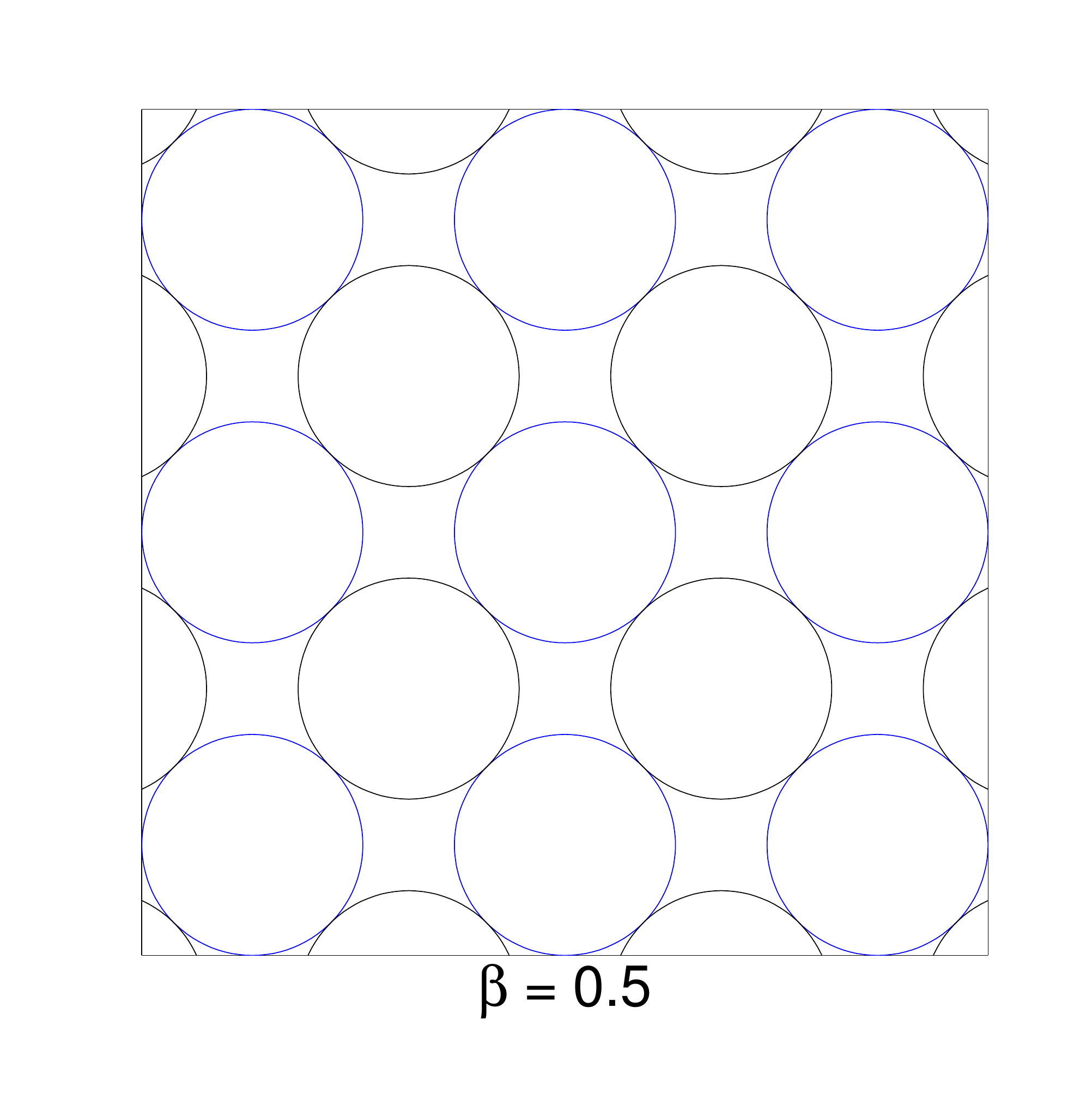}\hspace{-0.1cm}\includegraphics[width=3.1cm]{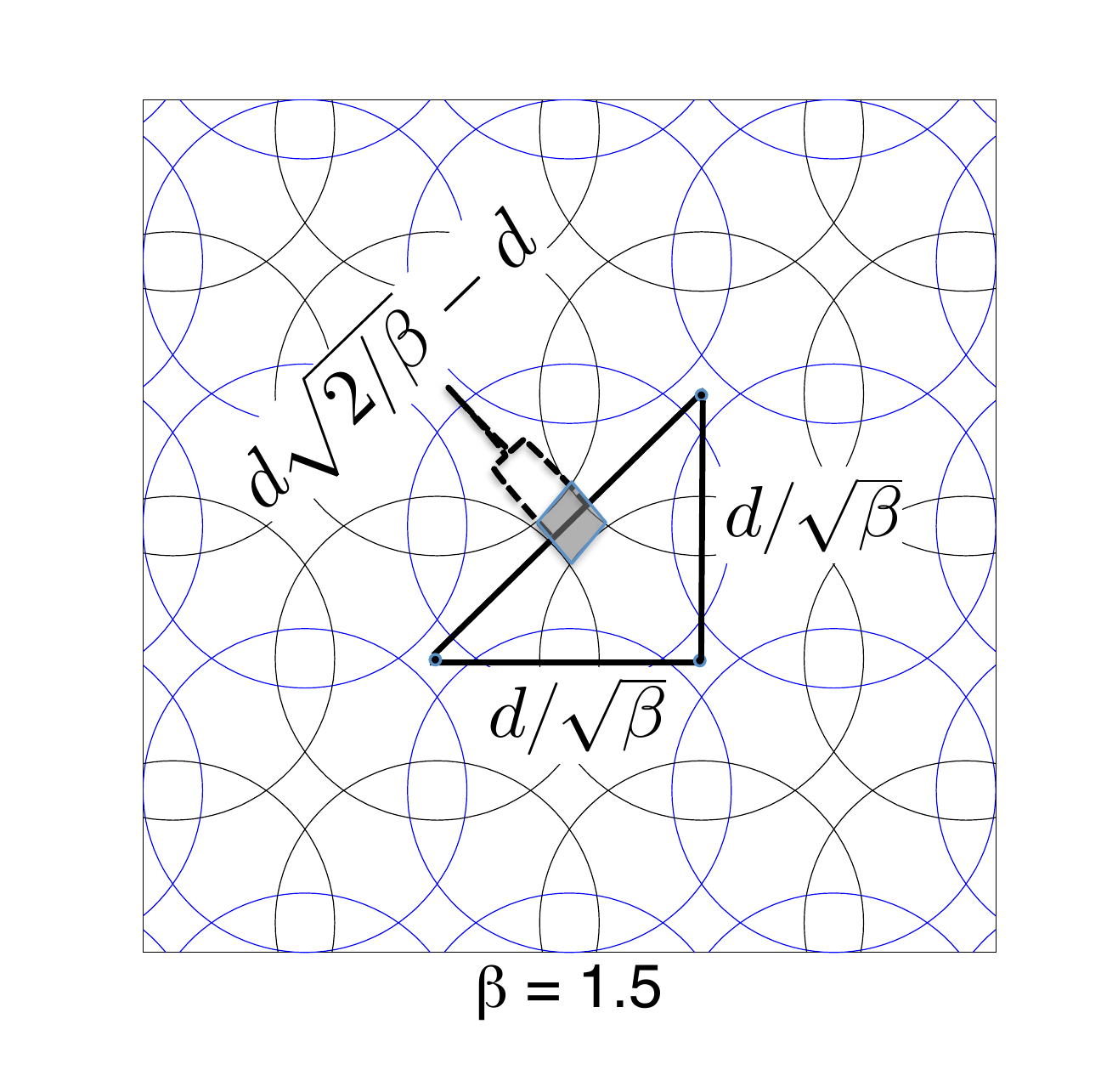}\hspace{-0.1cm}\includegraphics[width=3.1cm]{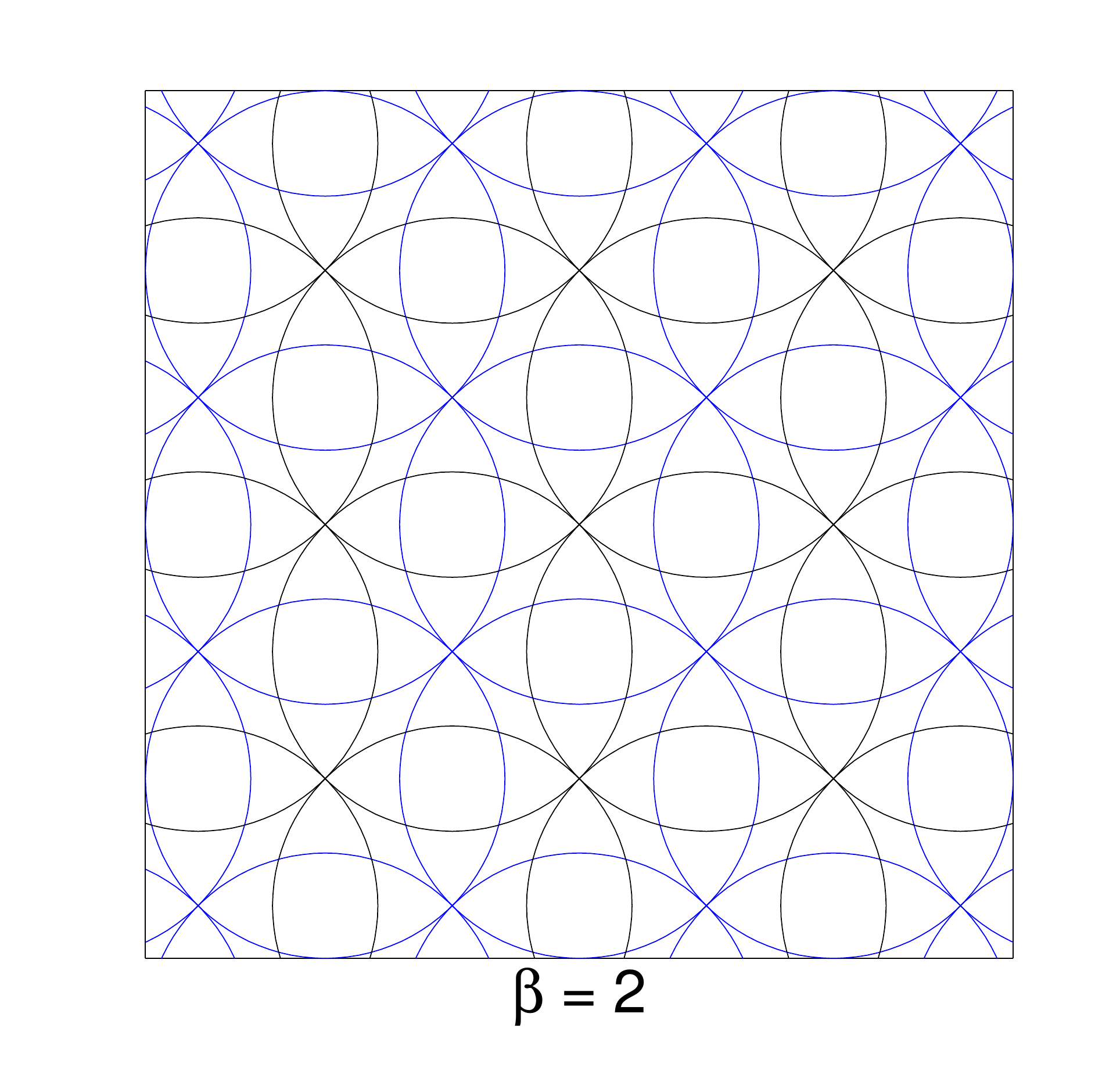}}
\vspace*{-.2in}
\caption{Lattice based user scheduling with $c = 3$}
\label{fig_beta15}
%\vspace*{-.2in}
\end{figure}

As explained next, the maximum packing can be obtained by putting the discs on a hexagonal lattice. In particular, consider a hexagonal lattice in the form of two sets of offset square-grid sub-lattices. Letting $d = 2r_0$ denote the diameter of the user discs with area of size  $D= (\pi/4) d^2$, we consider spacing the lattice points at a distance of $d/\sqrt{\beta}$. Such lattice examples for various values of $ \beta \in \{ 0.5,1.5,2\}$ are shown in Fig. \ref{fig_beta15} where blue and black circles simply correspond to the two set of discs on the two square sub-lattices. To avoid ``edge" effects we scale the area $\Ac$ to ``match" the lattice. Assume that $c^2$ discs are spaced on each of the blue and black rectangular sub-lattice (See Fig. \ref{fig_beta15} with $c = 3$ examples) and that there is one scheduled user at the center of each disc.  With this lattice-based scheduling there are $K_{\Lc}(\beta) = 2 c^2$ many scheduled users in each slot. For a given $ \beta$, $A = c^2 d^2 /\beta$ and the set of active users is $K_{\Lc}(\beta) = (\pi/2)\beta( A/D)$.  As the figure illustrates, for $\beta<2$, all active users are served as there is at least a point within each active users disc  that is not overlapped by  other user discs. Since for $\beta>2$, no scheduled user can be served (as any point in its region  is covered by other user discs), the multiplexing gains per dimension are maximized with $\beta=2$ yielding an upper bound on the multiplexing gain equal to $m_{\rm max}= \pi A/D$.

Next consider lattice-based scheduling in the case of  finite $N$, there is a trade-off between the number of active users and the probability that a user is served. At one extreme of $ \beta = 0.5$,  scheduled users are so sparsely located that active user discs do not overlap. In this case as long as an RRH site falls within a user's disc, any scheduled user is served. As we increase $ \beta$ from 0.5 to 2 the area where the BS has to fall for the user to be served (e.g., the gray area in Fig.\ref{fig_beta15}-$\beta = 1.5$) shrinks and eventually becomes a single point at $ \beta = 2$. Clearly, as $ \beta$  is increased, more users are scheduled but the probability that a user can be served becomes smaller. 

The maximum multiplexing gain  per RE in a lattice based user scheduling for finite $N$ is given by: 
\begin{equation}
\label{eq_m_LU_star}
m^{*\rm LU}_{1}(N) =  \max_{\beta} K(\beta)  p_1(\beta,N), 
\end{equation}
where $p_1(\beta,N)$ is the probability that at least one RRH can serve a user assuming a scheduling lattice with spacing $d/\sqrt{\beta}$. An RRH site can serve a user if it falls in the region of the user disc where other user discs overlap. Letting $\lambda(\beta)$ denote the area  of this  region, the probability of an RRH site fall within this region is given by $p_1(\beta,N) =  1 - [1-\lambda(\beta)/A]^{N}$.

As seen in Fig. \ref{fig_beta15}, for large values of $\beta$, we can approximate this region by the smallest square that includes that area.The square has sides equal to $ d\sqrt{2/\beta}-d$, so the area of interest is given by $ \lambda(\beta) \approx d^2 (\sqrt{2/\beta}-1)^2 = (4/\pi) D (\sqrt{2/\beta}-1)^2 $. 
%As the above derivation is valid for any disjoint group, $m^{*total,\rm LU}_{Q,1}(N)  = Qm^{*\rm LU}_{1}(N) $.  Curve $m^{*total,\rm LU}_{Q,1}(N) $ in Fig. \ref{fig_MG_vs_N_curves} is calculated substituting (\ref{eq_p1}) into (\ref{eq_m_LU_star}).

\subsection{Random Scheduling Simulations}
\begin{figure}
\centering
\centerline{\includegraphics[width=7.5cm]{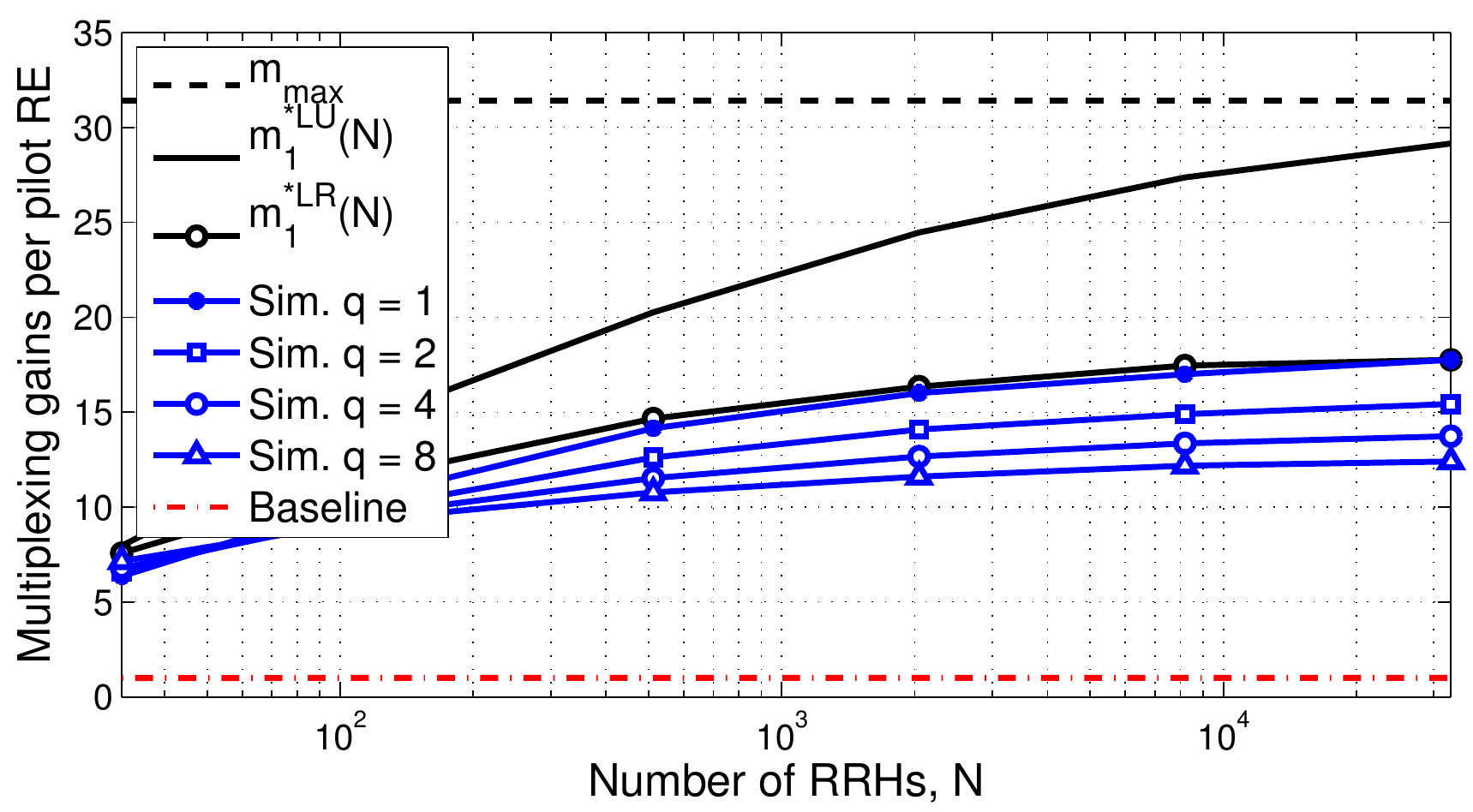}}
\vspace*{-.1in}
\caption{$A/D = 10$ and $Q = 8$}
\label{fig_MG_vs_N_curves}
\vspace*{-.2in}
\end{figure}

Fig. \ref{fig_MG_vs_N_curves} compares  the upper bound $m_{\rm max} $, $m^{*\rm LU}_{1}(N) $ and the performance of the baseline scheme where  only 1 user per RE per RB is served by the system. For high enough number of RRHs where each user has at least one RRH in its vicinity, the baseline scheme has an average multiplexing gain of $1$ per pilot RE. Besides, for various $q$ values, $100$ frames with random user scheduling and random RRH site locations  are run for each $K$ and $N$. The simulated multiplexing gain per pilot RE $m_q(K,N)$ and $m_q^{*}(N)$ (as in (\ref{mq}) \& (\ref{mq-opt})) are obtained. 

We first focus on the $q = 1$ case. As expected, $m_{\rm max} $ is an upper bound to both $m_1^{*}(N)$ (Sim. $q = 1$ curve) and  $m^{*\rm LU}_{1}(N) $ (lattice based scheduling). It can be seen that as the number of RRH, $N$, increases the ratio of  $m_{\rm max}$ to $m_1^{*}(N)$, empirically converges to $\pi/2$. We can also observe that $m^{*\rm LU}_{1}(N)$ approaches $m_{\rm max} $ as $N$ increases and the lattice based scheduling has better performance than random scheduling. 
Fig.  \ref{fig_MG_vs_N_curves} also shows  $m^{*\rm LR}_{1}(N) $, the multiplexing gains per dimension in the case where the RRH sites are placed one a lattice (similar to the earlier described user lattice) with random UEs scheduling. As seen,  the benefits from careful placement of the RRH sites are only marginal with respect to random RRH site placement.

 Fig. \ref{fig_MG_vs_N_curves}  also shows the multiplexing gains for random UE scheduling and random placement of the RRH sites, with $q=2$, 4, and 8. As seen, aligning user codes in a single pilot dimension (i.e., $q = 1$) yields higher multiplexing gains per RE than the $q=2$, 4, or 8. 

\begin{figure}
\centering
\centerline{\includegraphics[width=7cm]{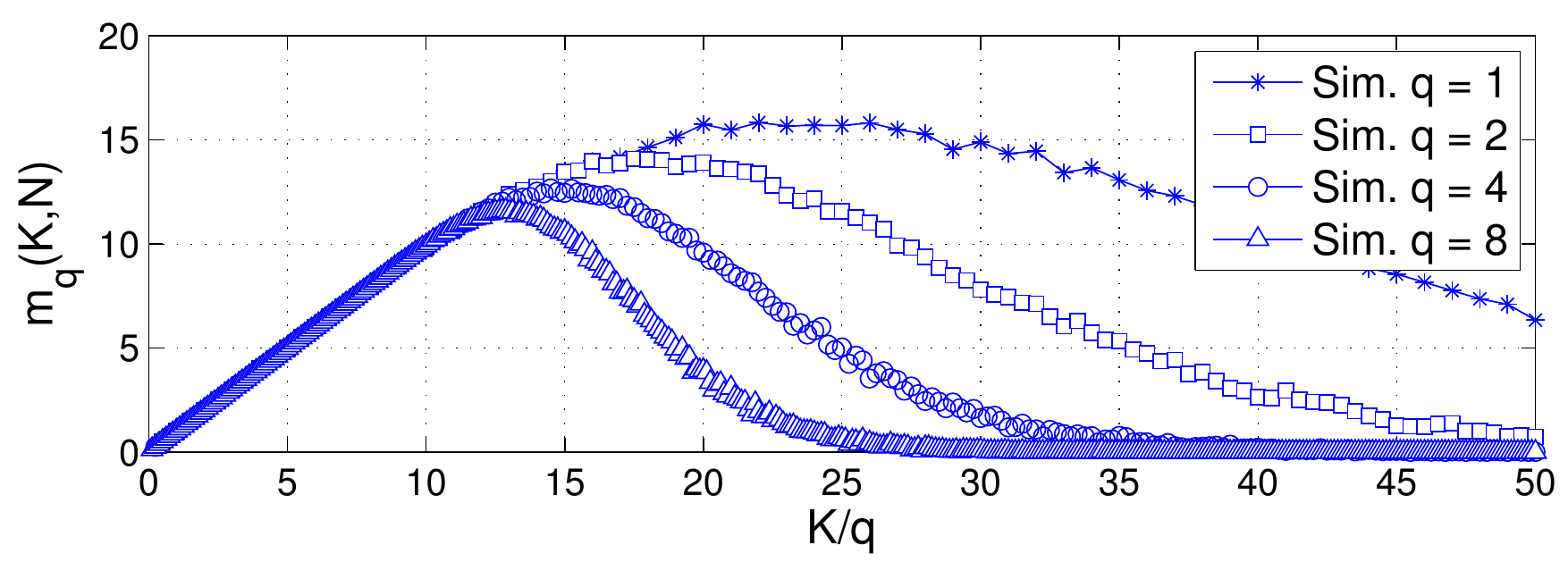}}
%\vspace*{-.13in}
\caption{$N$ = 2048}
\label{fig_MG_vs_NK_curves}
%\vspace*{-.1in}
\end{figure}

\begin{figure}
%\vspace*{-.1in}
\centering
\centerline{\includegraphics[width=7cm]{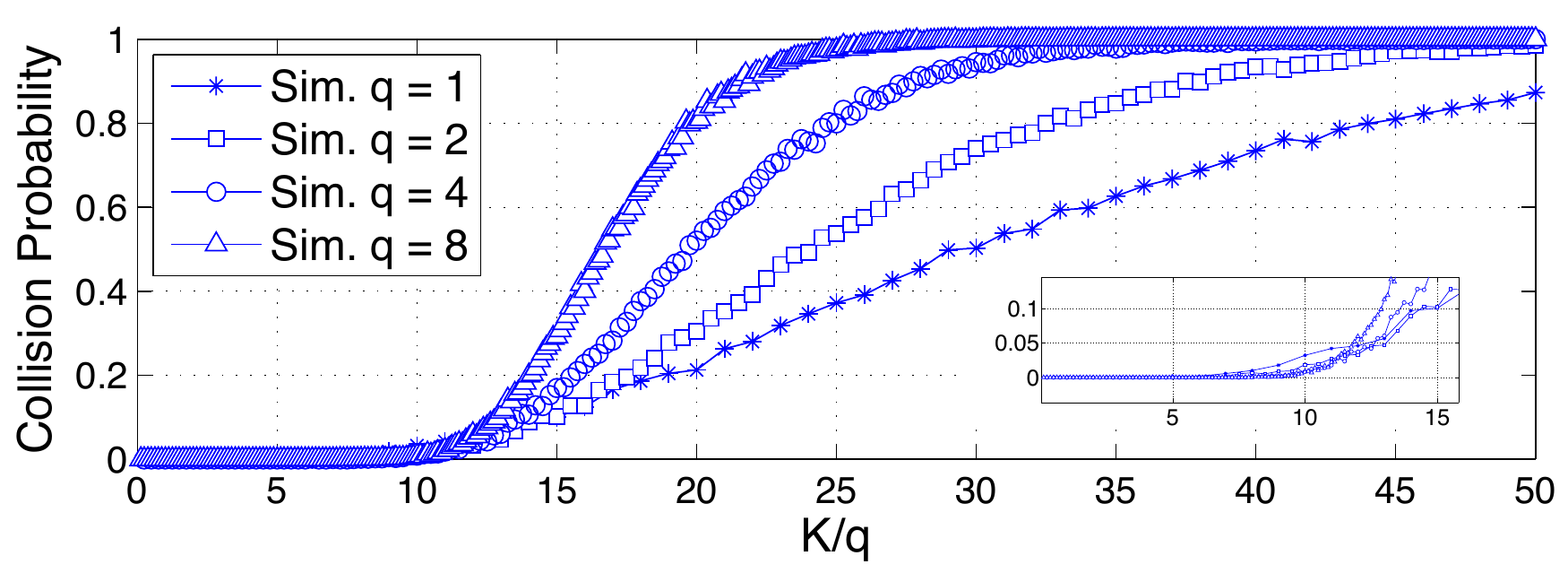}}
%vspace*{-.13in}
\caption{$N$ = 2048}
\label{fig_MG_vs_collision_prov}
%\vspace*{-.2in}
\end{figure}

Figs.~\ref{fig_MG_vs_NK_curves} and \ref{fig_MG_vs_collision_prov} shed some light into why $q=1$ performs best. 
Fig.~\ref{fig_MG_vs_NK_curves} shows the multiplexing gains per RE  as a function of the number of scheduled users per RE. Inspection reveals that the optimal number of active-users per dimension, $K^*_q(N)/q$,  decreases with increasing $q$ values. Fig.~\ref{fig_MG_vs_collision_prov} shows the active-user collision probability (see (\ref{Sjtq})), as a function of the number of scheduled users per RE, and provides some insight into the trend observed in Fig.~\ref{fig_MG_vs_NK_curves}. While at small numbers of active users per RE, the collision probability is lower at larger $q$ values (see figure inset), in the performance-optimizing regime of large numbers of active-users per dimension,  the collision probability is much lower for $q=1$.  To further understand this, consider a system with $Q>1$ pilot REs per RB,  with $q=1$ and $q=Q$. For $q=Q$ RRH $j$ serves no user  if $|\Kc_j(t)|$ exceeds $Q$.  In contrast, with $q=1$, some users may be served even when  $|\Kc_j(t)|$ exceeds $Q$, as a user is only interfered by the subset of users sharing the same pilot dimension. Effectively, the benefits of the $q = 1$ system can be attributed to ``pilot interference alignment'' of the other-group user pilots away from the user's direction. 
%\vspace{-0.2cm}

\subsection{Finite Angular Spreads}

Although aggressive pilot reuse and fast user proximity detection can offer substantial increases in multiplexing gains with respect to conventional pilot-assignment schemes, these gains come at a large cost in the number of RRH-sites required. As the earlier examples reveal, the increase in multiplexing gains is sub-logarithimic in the number of RRH sites required.  In this section we leverage the presence of antenna arrays at each RRH site to improve  the number of RRH sites vs. multiplexing gain trade-offs. We remark that placing many array elements on a small footprint at each RRH unit becomes increasingly feasible at higher (e.g., mmWave) frequencies and allows RRH-site sectorization.  Sectorization is a well known technique that increases the spectral efficiency per site in cellular networks by partitioning each site radially into sectors and reusing the spectral resources in each sector\cite{valenzuela-sector}.   

We assume a simplified scenario where the  channel of any given user  in the proximity of a given RRH site (i.e. within distance $r_o$) has a finite angular spread $\theta>0$.  An RRH can separate the received pilot observations (\ref{Ymj-pilots}) into angular ``sectors'' by appropriate  spatial filtering on $\Ym_j[n]$ (for a given sector this may correspond to, e.g.,  projecting $\Ym_j[n]$ onto a set of DFT vectors spanning a sector's angular frequency range).  

Note that an active user in the proximity of an RRH site (e.g. within distance $r_o$) will  appear to be present (i.e., its pilot will be received at sufficiently high power) on only the subset of the RRH site sectors that have (significant) overlap with the user's angular spectrum support. Consequently we assume that user $k$ is in proximity of a sector $s$ of RRH $j$, if the distance between user $k$ and RRH $j$ is less than $r_o$, {\em and} if the intersection of the supports of the angular spectrum of user $k$ and  sector $s$ of RRH $j$ is non-empty.  We can thus consider straighforward extensions of the techniques of the preceding section replacing the notion of RRH sites with RRH-site sectors. For instance, in the system with $q=Q=1$, an RRH-site sector can serve an active user if it is the only active user in proximity to the RRH-site sector. 

\begin{figure}
\centering
\centerline{\includegraphics[width=6cm]{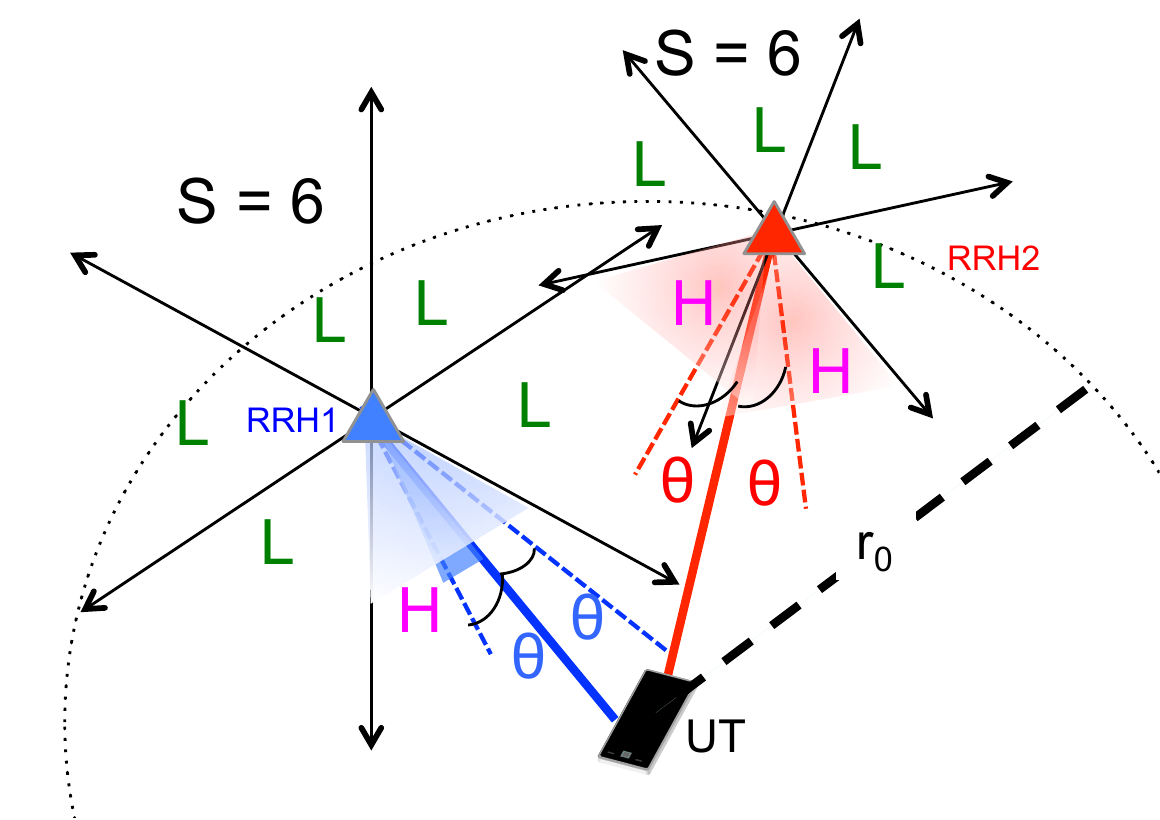}}
%\vspace*{-.1in}
\caption{Two RRHs are shown within $r_0$ distance from UT. UT is in the proximity of a sector if its angular spectrum support overlaps with the sector. The sectors where the UT is in the proximity are denoted by $\sf H$, the others are denoted by $\sf L$. }
\label{fig_sectors}
%\vspace*{-.2in}
\end{figure}

In Fig. \ref{fig_sectors}, a sectorization abstraction is shown where we plot two RRHs within the proximity of one UT with angular spread $\theta$. Each RRH site has $6$ sectors shown by regions between two consecutive arrows. In this figure,  the UT is in the proximity of one of the sectors of RRH 1 while it is in the proximity of two sectors of RRH 2 as shown by letter $\sf H$.

\begin{figure}
\centering
\centerline{\includegraphics[width=7.5cm]{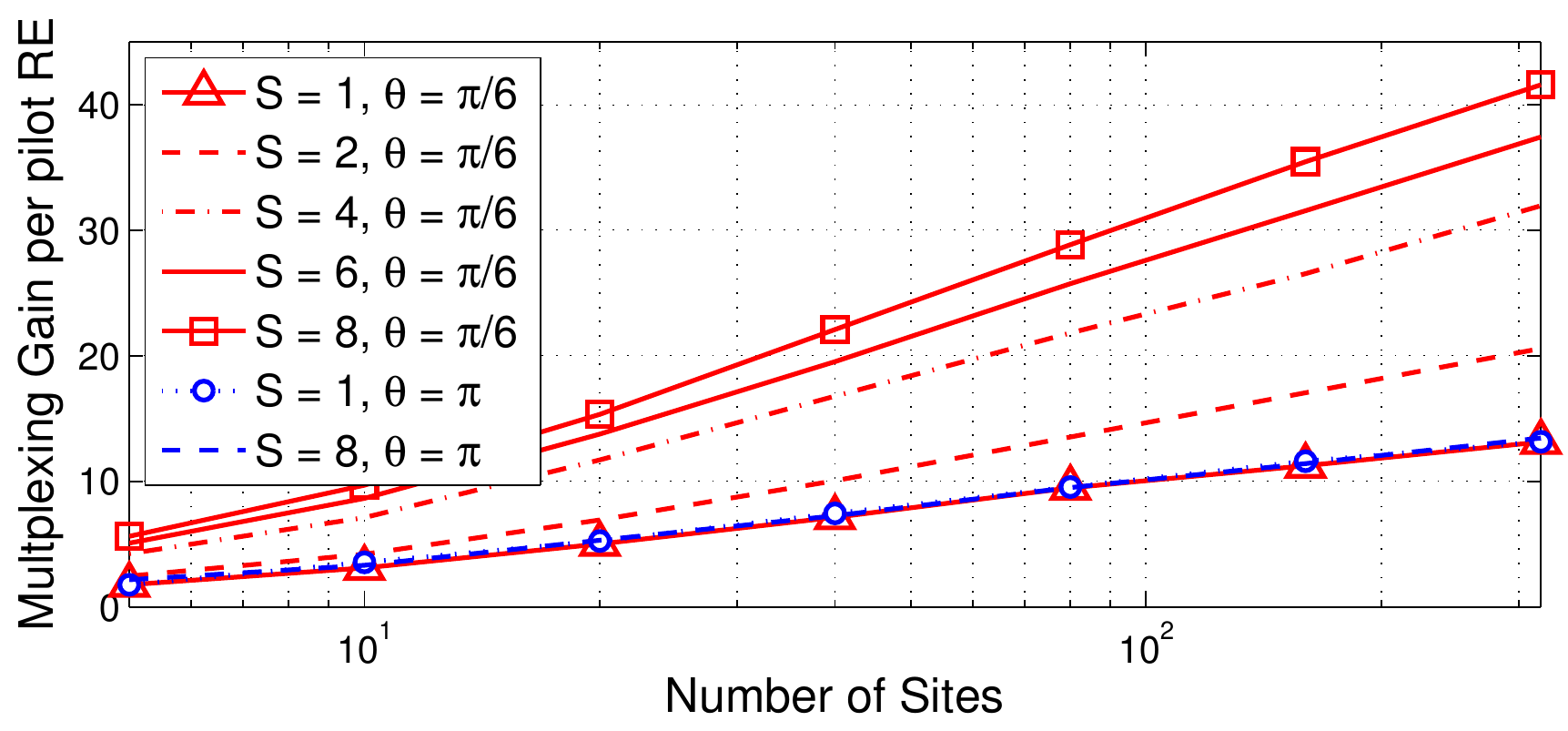}}
%\vspace*{-.1in}
\caption{$q = 1, Q = 8$}
\label{fig_Sector_vs_NumSites}
%\vspace*{-.2in}
\end{figure}

%Assuming zero angular spread ($\theta$), a user within the vicinity of RRH site can transmit to and receive from this RRH-site on a particular sector, if the LOS direction between the user and RRH-site is within the limits of this sector. Increased number of sectors can help avoiding collisions from nearby users by possibly putting them into different sectors which in turn can increase multiplexing gains. When angular spread is nonzero, $\theta>0$, sector limits should be compared with a range of directions obtained by tilting the LOS direction with $\pm \theta$. In this case, a user can be served in more than one sector. Actually, at an extreme case when $\theta = \pi$, any site, possibly with more than one sector becomes effectively an omnidirectional, single sector site. 
Fig. \ref{fig_Sector_vs_NumSites} illustrates the benefits of sectorization comparing against the omni-scenario in Fig. \ref{fig_MG_vs_N_curves} with $q = 1$ and $Q = 8$. The figure considers user angular spreads of $\theta=\pi$ (as in Fig. \ref{fig_MG_vs_N_curves}) and $\theta = \pi/6$. As expected, for $\theta=\pi$ we get the ``omni'' performance in Fig. \ref{fig_MG_vs_N_curves}  with $S = 1$ and  $S = 8$ sectors. When the user-angular spread, however, is $\theta = \pi/6$, sectorization provides substantial gains. Indeed, the multiplexing gain obtained by $10^4$ RRHs  (see Fig. \ref{fig_MG_vs_N_curves}) can be obtained by $45$ RRH sites if $S = 4$ sectors are used,  by $30$ sites if $S = 6$ and by $23$ sites if $S = 8$.

In this section, we exploited the narrow angular spread of user channels by considering sectorization. In \cite{Gesbert-aoa}, the same characteristic of user channels is used to carefully design user schedules. In contrast, here, RRH-sector proximity detection combined with aggressive pilot reuse allows us to randomly schedule users while still maintaining high multiplexing gains.

%{\RED Add citation for sectors to the bib file, , finite angular spread citation}

%@INPROCEEDINGS{5757638, 
%author={Huang, H. and Alrabadi, O. and Daly, J. and Samardzija, D. and Tran, C. and Valenzuela, R. and Walker, S.}, 
%booktitle={Signals, Systems and Computers (ASILOMAR), 2010 Conference Record of the Forty Fourth Asilomar Conference on}, 
%title={Increasing throughput in cellular networks with higher-order sectorization}, 
%year={2010}, 
%pages={630-635}, 
%keywords={MIMO communication;antenna arrays;array signal processing;cellular radio;anechoic chamber;cellular networks;circular antenna array;fixed beamforming;higher-order sectorization;multiuser MIMO transmission;single-antenna transmission;Antenna arrays;Arrays;Geometry;Interference;Signal to noise ratio;Throughput}, 
%doi={10.1109/ACSSC.2010.5757638}, 
%ISSN={1058-6393}, 
%month={Nov},}

\section{Pilot Coding for Fast Proximity Detection}
\label{sec_codes}
In this section we present codes for active-user proximity detection.  Following the approach in \cite{marzetta-massive}  each scheduling slot spans the whole bandwidth and comprises the totality of a set of consecutive OFDM symbols. The time-duration of a slot is within the coherence time of the user channels and that  the maximum user-channel multipath spread is $L$ samples long (with $L$ not exceeding the OFDM circular prefix). $L$ pilot dimensions per user (on distinct OFDM tones) are needed to learn a user's channel over the whole bandwidth for the duration of such a scheduling slot\footnote{In practice, $\bar L =  L+\Delta,\Delta>0$ many pilot dimensions can be used to ensure the quality of channel estimation based on any $\bar L$ random pilot locations over OFDM block. In this case, the analysis provided in this section will be valid by using $\bar L$ instead of $L$.}. In terms of the required training overheads to learn a user's channel, this setting is equivalent  to the abstracted scenario in the previous section whereby each scheduling slot comprises $L$ concurrent RBs and the user channels are quasistatic over each RB \cite{marzetta-massive}.  Furthermore, the $Q$ pilot dimensions per RB (of the previous section) correspond here to assuming that within each scheduling slot a set of $QL$ orthogonal pilot vectors (spanning $QL$ OFDM time-frequency REs) are allocated for UL training.

\begin{figure}
\centering
\centerline{\includegraphics[width=7cm]{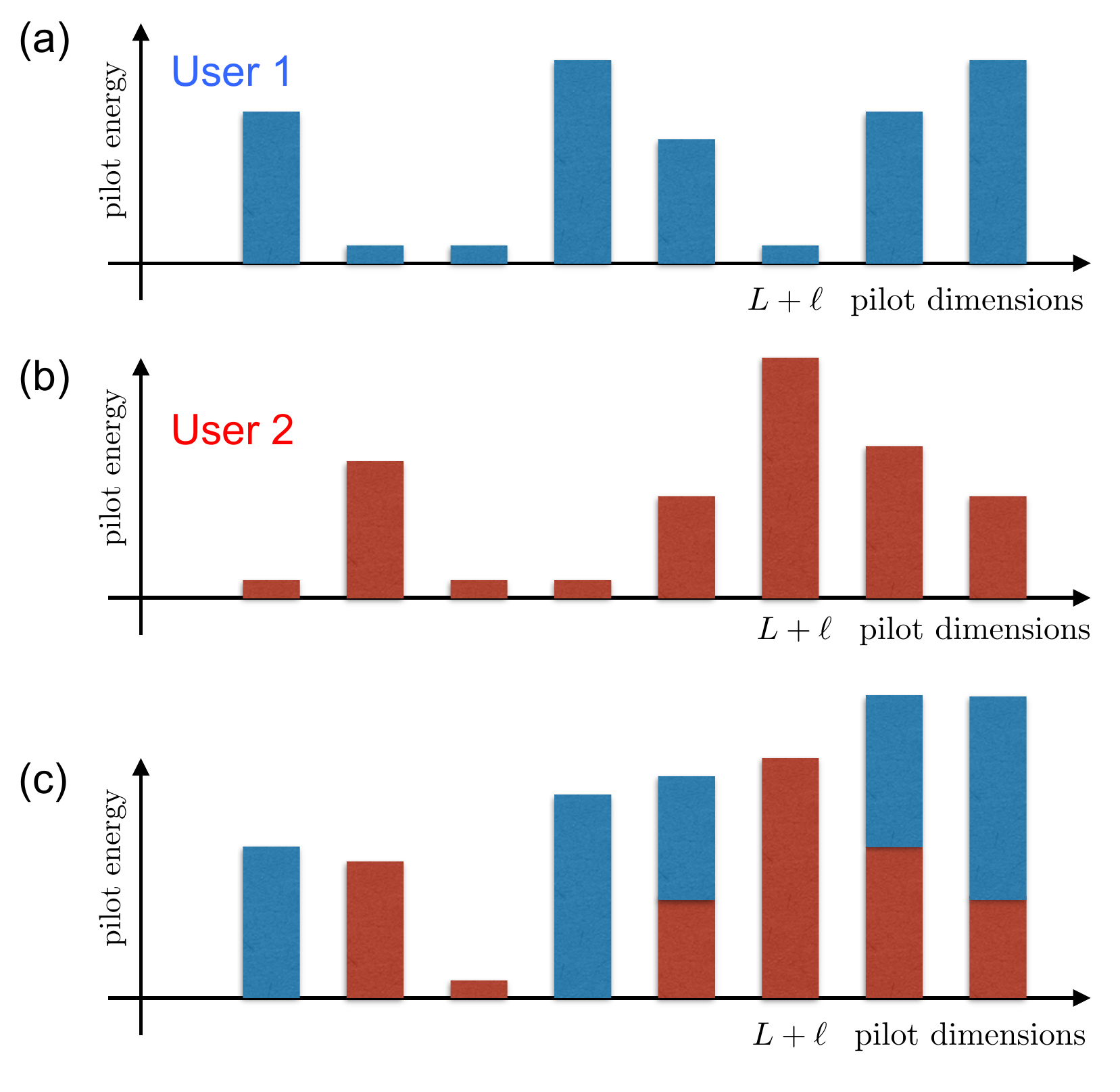}}
%\vspace*{-.30in}
\caption{Received pilot energy}
\label{fig_code_pilots}
%\vspace*{-.2in}
\end{figure}

Since at least $L$ UL pilot dimensions are required for learning a users' channel, we consider the case whereby a set of $L'=L+\ell>L$ pilot dimensions (for some $\ell>0$ to be determined) are aggressively assigned to a set of $K$ active users across the RRH coverage area. Without loss of generality, we assume that these pilot dimensions correspond to a set of $L'$ REs (on distinct tones) in the OFDM plane. We enumerate the pilot REs shared by an active group from 1 to $L'$ and consider ``on-off'' type pilot codes. The $k$-th active user pilot pattern  is specified by means of an $L'\times 1$ binary vector $\bv_k$, describing whether or not user $k$ transmits a pilot in each of the $L'$ RBs in the scheduling slot. Let $b_k[n] = \{\bv_k\}_n$, then user $k$ transmits  a pilot on shared pilot RE $n$ if  $b_k[n] =1$ and remains silent if $b_k[n]=0$.  In Fig.~\ref{fig_code_pilots}, a simple example is shown with $L = 5$ and $\ell = 3$ where two users share $8$ pilot dimensions.  

Fig.~\ref{fig_code_pilots} (a) shows the received pilot energy at an RRH if only the first user is in the proximity of this RRH while (b) shows the received pilot energy if only the second user is in the proximity of this RRH. The probability of any pilot dimension being in deep fade is negligible due to large $M$ and coherent combining, and noise floor is easily distinguished from any received pilot energy. Then the individual received pilot energy plots (Fig.~\ref{fig_code_pilots} (a) \& (b)) can be also seen as a visualization of the on-off pilot pattern for each user. The specific on-off code assignment to each user in this example lets two users' pilots overlap at pilot dimensions $5, 7$ and $8$. The received pilot energy at a nearby RRH (within $r_0$ distance to both users) is the superimposition of two pilot sequences, shown in Fig.~\ref{fig_code_pilots}~(c).

Next we consider user proximity detection at a fixed RRH site and suppressing the dependence of variables on RRH site index. Let $z_k=1$ if user $k$ is within distance $r_o$ of RRH $j$, and $z_k=0$  otherwise. 
According to the example in Fig.~\ref{fig_rrh_reqs}, to enable proximity detection,  the detection mechanism based on a set of active-user codewords must satisfy the following:
\begin{itemize}
\item if multiple users are within distance $r_o$ of the RRH site (i.e., if $\sum_{k=1}^K z_k>1$), the RRH must be able to determine that there is a pilot collision;% in the given pilot dimension;
\item if a single active user is within distance $r_o$ of the RRH site (i.e., if $\sum_{k=1}^K z_k=1$), the RRH must be able to identify that a single user is in proximity and the identity of that user  (i.e., the $k$ index for which $z_k=1$);
\item the RRH must also be able to identify when no users are in proximity of the RRH.
\end{itemize}
We also note that if a single active user on these $L'$ pilot dimensions is the proximity of the RRH and is detected by the RRH, the user can be served by the RRH provided the user channel can be estimated, that is, provided the user has transmitted  a pilot over at least  $L$ out of of  $L'$ shared pilot REs (i.e., the user codeword must have at least  $L$ ones). Note that, the code example in Fig.~\ref{fig_code_pilots} with two users satisfies these conditions. If the RRH observes higher than noise floor energy at more than $3$ locations, it rightfully declares a collision (the case seen in Fig.~\ref{fig_code_pilots} (c)). If it observes exactly $3$ ``off" pilot dimensions, by matching the on-off pattern it decides who the unique user is (Fig.~\ref{fig_code_pilots} (a) or (b)). In case it observes no received pilot energy in any of the pilot dimensions, it declares there are no users in its proximity signaling at these pilot dimensions.

Inspired by the received pilot energy example shown in Fig.~\ref{fig_code_pilots}, the pilot energy detection at an RRH can be seen ``OR"-type channel  (formal justification is also provided at the end of the section),  whereby an RRH receives an ``$1$'' (indicating sufficiently high received power) if at least one active user transmitting a pilot is in the proximity of the RRH and $0$ otherwise. Specifically,  for all $1\le n \le L'$, the RRH at pilot RE $n$ observes the following:
 \begin{equation}
 \label{or-channel-eqn}
 \epsilon[n]  = {\rm OR}(z_1 b_1[n], z_2 b_2[n], \cdots, z_K b_K[n]) \ , 
 \end{equation}

The simplest codes that enable active-user proximity detection are comprised of $K\le L+1$ codewords (corresponding to the case $\ell=1$) given by
\begin{equation}
\label{max-eff-codes}
b^{(1)}_k[n] = 1 -\delta[k-n], \;\text {for $1\le k\le K$.}
\end{equation}

It can be verified that for the user-proximity  model (\ref{or-channel-eqn}), the observations $\{\epsilon[n]; \ \  1\le n \le L'\}$ satisfy the following:
\begin{equation*}
 \epsilon[n]   =\begin{cases} 1 & \text{if $\sum_{k=1}^K z_k>1$} \\ 1\!-\!\delta[n\!-\!k_o] & \text{if $z_k=\delta[k\!-\!k_o]$ for some $k_o$} \\ 0 & \text{if $z_k=0$ for all $1\le k \le K$} \end{cases}.
\end{equation*}

Consequently, if the RRH receives the all 1's pattern (active-user pilot collision) or the all 0's pattern (no active user is close by) it does not send any user data. If, however, it receives a pattern $\epsilon[n] = 1-\delta[n-k_o]$, it can identify the single user in proximity as user $k_o$.  Effectively, a single user is present when there is exactly one zero observed, and the index of the pilot RE where a  zero is observed  identifies the user  in proximity (as this is the only user that did not transmit a pilot on the given pilot RE). Subsequently, when user $k_o$ is identified as the single user in proximity, the set of $L$ pilot observations on the $L$ pilot REs except  pilot RE $k_o$ allow the RRH to estimate the user channel across the whole bandwidth and thus serve the user in the data portion of the scheduling slot.

Since $L+1$ pilot REs are used per user, as opposed to the minimum required of $L$, the pilot code efficiency is $\eta= L/(L+1)$. Furthermore,  letting $K_{\rm max}$ denote the number of code codewords, the maximum number of active users that can be supported on the common set of $L'$ pilot REs by a given code  is $K_{\rm max}$.  For the code  in (\ref{max-eff-codes}), $ K_{\rm max}=L+1$ users.

Extensions of the code in (\ref{max-eff-codes}) can be developed that trade off  $\eta$ with  $ K_{\rm max}$. One such family of codes that includes the code in (\ref{max-eff-codes}) is parametrized by a pair of integers $L$ and $\ell$ with  $\ell\ge 1$. The code for a given $(\ell, L)$ pair is the constant weight code comprising all  binary codewords of length $L'=L+\ell$, with $L$ ones and $\ell$ zeros. The active users using such a code share $L'=L+\ell$ REs for UL training.

Consider using such a code for a fixed $\ell$  and assume each active user (sharing the $L+\ell$ REs for UL training) is assigned a unique codeword. For the model (\ref{or-channel-eqn}), it can be readily verified that if more than 1 active users are in the proximity of the RRH, then there are at most $\ell-1$ zeros in $\{\epsilon[n]\}$, while the presence and identity of a single user in proximity are readily recovered at the RRH from the set of $\ell$ values of $n$ for which $\epsilon[n]=0$. Also, the observations on the $L$ pilot REs where the detected user has ones in its codeword allow the RRH to estimate the active user channel over the whole bandwidth and serve the user. Clearly, $K^{(\ell)}_{\rm max} = {L+\ell \choose \ell}$,  and $\eta^{(\ell)} = L/(L+\ell)$.

Given a  target value for $K$, the number of active users on a set of pilot REs, we may select the code
(among the ones for which $K^{(\ell)}_{\rm max}\ge K $) that yields the highest efficiency. This is equivalent to finding the lowest $\ell$  for which   $K\le {L+\ell \choose \ell}$. Hence, the highest efficiency for a given $K$ is given by

\vspace{-0.5cm}
\begin{equation*}
\label{max-code-eff-vs-K}
\eta^*\!(K;\!L)\!  =\!
\begin{cases}
1 \!&\!\! \text{if $K=1$}\\
\frac{L}{L+\ell} \!&\! \!\text{if ${L\!+\!\ell\!-\!1 \choose \ell-1}\!< \!K\!\le\!{L\!+\!\ell \choose \ell}$ for some $ \ell\!\ge \!1$}
\end{cases}
\end{equation*}
%\vspace{-0.5cm}

Subsequently, the achieved net multiplexing gains by the RRH system is given by $m_{\rm net} (K,L) = m(K) \,\eta^*(K;L) \ .$

\begin{figure}
\centering
\centerline{\includegraphics[width=7cm]{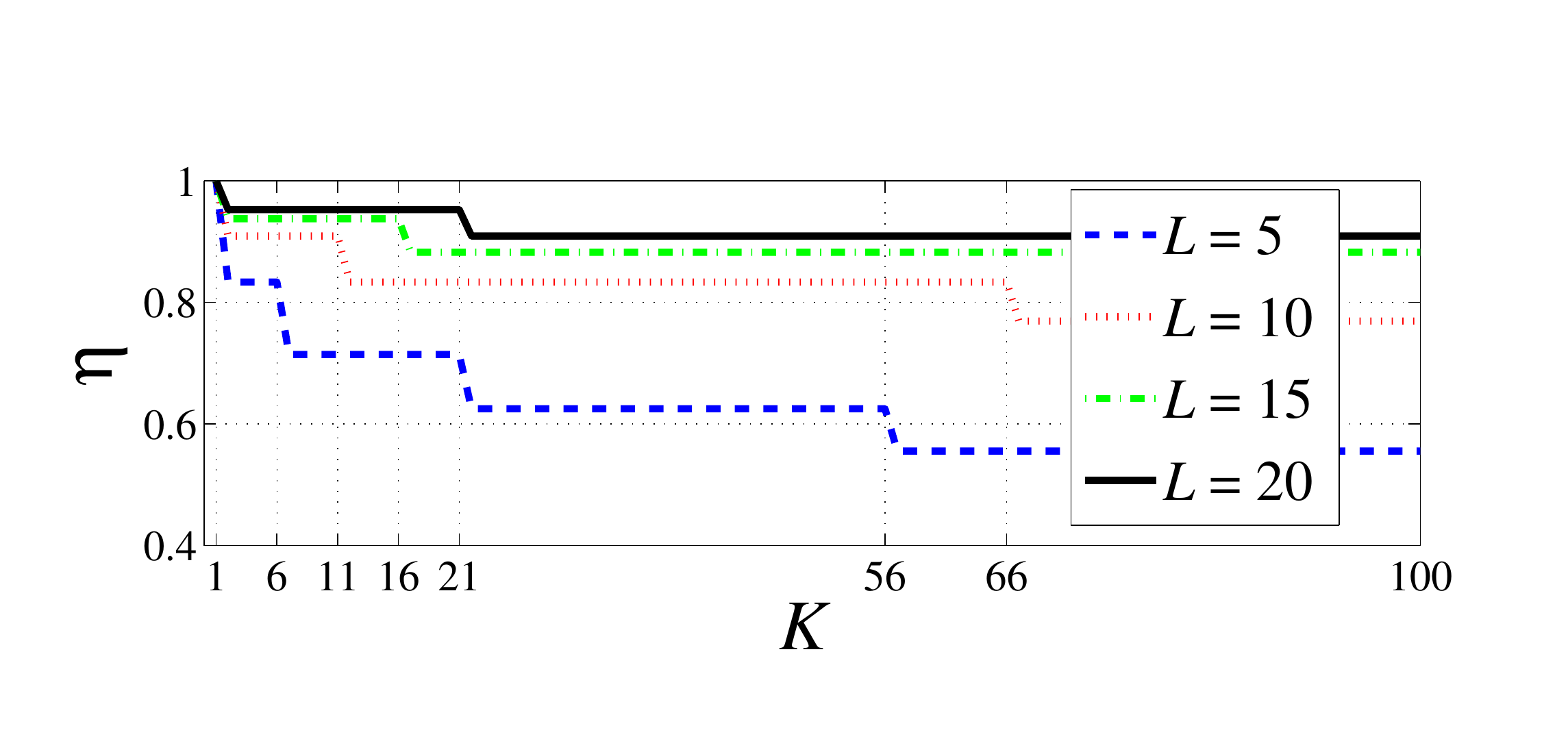}}
\vspace*{-.30in}
\caption{Code Efficiency vs. number of active users per pilot dimension for various values of $L$ .}
\label{fig_code_eff}
\vspace*{-.2in}
\end{figure}

Fig.~\ref{fig_code_eff} shows the maximum efficiency possible with the given family of codes as a function of $K$, for various values of $L$.  As seen, even small values of $L$ provides high efficiencies.
\vspace{-0.04cm}
Finally, it is worth justifying the use of the OR channel in (\ref{or-channel-eqn}) at  RRH site $j$.   Given an $L$-tap channel $\tilde{\hv}_{k}[\tau]$ between user $k$ and  RRH site $j$ (suppressing again the dependence of variables on $j$), the channel response on  tone $n$ is given by
\[
\hv_{k}[n] = \sum_{\tau=0}^{L-1} \tilde{\hv}_{k}[\tau]e^{-j\frac{2\pi}{N}n\tau}
\]
where $2\pi/N$ is the OFDM tone spacing.
Assuming also that the $\tilde{\hv}_{k}[\tau]$'s are independent in $k$ and $\tau$, and that $\tilde{\hv}_{k}[\tau]\sim\Cc\Nc(0, \rho_{k,\tau}\Id)$ with $\rho_{k,\tau}$ unknown we have $g_{k} = \sum_{\tau=0}^{L-1} \rho_{k,\tau}$.

Next, note that the observation on the $n$-th pilot RE ($n$-th OFDM tone) is given by the $Q=1$ specialization of (\ref{Ymj-pilots})
\begin{equation*}
\label{yvj-pilots}
\yv[n] = \sum_{k=1}^{K} b_{k}[n] \hv^\transp_{k}[n]  +\wv[n]
\end{equation*}
with $b_k[n]$ denoting the pilot of user $k$ on the $n$-th RE (with $b_k[n]\in\{0,1\}$).   The RRH site first obtains the sample-average received energy per antenna estimate $\hat{\Ec}[n] =\| \yv[n] \|^2/M$.

Noting that $\EE\left[\hat{\Ec}[n]\right] = \sum_{k=1}^{K} b_k[n] g_{k} +N_o$,
a hypothesis test of the form
\begin{equation*}
\hat{\Ec}[n]
\quad\mathop{\gtreqless}_{\hat{\epsilon}[n]=0}^{\hat{\epsilon}[n]=1}\quad \Gamma
\end{equation*}
for some appropriately defined threshold enables proximity detection. For the abstracted example of Sec.~\ref{sec_system_model},  where a user's large scale gain $g_{k}=z_{k}g$, i.e., it is a non-zero value $g$ if the user  $k$ is within $r_o$ distance of the RRH  $j$ and zero otherwise, setting the threshold to $\Gamma=0.5g+N_o$, and taking the limit $M\to\infty$ yields $\hat{\epsilon}[n] = \epsilon[n]$, with $\epsilon[n]$ from (\ref{or-channel-eqn}).

%\begin{figure}
%\centering
%\centerline{\includegraphics[width=3.5cm]{figures/coded-pilot.pdf}}
%\vspace*{-.1in}
%\caption{Superposition of pilot energy}
%\label{fig_code_eff}
%%\vspace*{-.2in}
%\end{figure}

\bibliographystyle{IEEEtran}
\bibliography{IEEEabrv,refs}

\end{document}